\documentstyle[12pt]{article}
\newcommand{\nl}{\nonumber \\}
\newcommand{\be}{\begin{equation}}
\newcommand{\ee}{\end{equation}}
\newcommand{\ba}{\begin{eqnarray}}
\newcommand{\ea}{\end{eqnarray}}
\newcommand{\ci}[1]{\cite{#1}}
\newcommand{\bi}[1]{\bibitem{#1}}
\newcommand{\la}[1]{\label{#1}}

\date{}
\begin{document}
\begin{center}
{\Large\bf Structural vertices of extended $SU(3)$-chiral lagrangians in the
large-$N_c$ approach}\footnote{Talk given at the Workshop on
Chiral Perturbation Theory and Other Effective Theories
(Karreb{\ae}ksminde, Denmark, Sept.1993)} \\

\medskip

{\large\bf A.A.Andrianov, V.A.Andrianov and V.L.Yudichev}\\
{\small\it Department of Theoretical Physics,
University of Sankt-Petersburg,
 198904 Sankt-Petersburg, Russia}

\end{center}

\bigskip

{\small We establish the connection between the structural coupling
constants of the phenomenological chiral lagrangian and the coefficients
of  effective lagrangians obtained in
the QCD-bosonization models by means of the derivative
expansion. The extension
of the chiral lagrangian describing matrix elements of the pseudoscalar gluon
density is examined. The large-$N_c$
relations for corresponding structural constants are elaborated. }

\section{Introduction}
\hspace*{3ex} The chiral lagrangians for light $SU(3)_f$-pseudoscalar mesons
\ci{We,GL}
describe their strong and electroweak interactions at low energies
with reasonable precision when the Chiral Perturbation Theory (CPTh) is
applied (see\ci{GL,Do,Ec} and references therein).
The modern chiral lagrangians reflect the exact, softly broken and
spontaneously broken symmetries of the Quantum Chromodynamics (QCD) and
include the interaction vertices compatible with these symmetries.
The corresponding coupling constants contain the detailed information about
the QCD dynamics and have not been yet calculated from the QCD
in a reliable way.
Practically they are established from the phenomenology \ci{GL,Do,Ec}.

On the other hand the QCD-bosonization approaches \ci{AA,DN} and the quark
models \ci{Ra,Ru,Vo}
have been developed to calculate
the interaction vertices of the $SU(3)$-chiral lagrangian. When following the
large-$N_c$ counting in the QCD one can derive the general form of the chiral
lagrangian which is characteristic for any bosonization model. This
"model-framework" lagrangian is obtained from the bosonization of
one-quark loop by application of the derivative expansion
and does not coincide with the
phenomenological one in some vertices.
Therefore it is of real interest to establish the
relations between them in order to reveal new constraints on the
phenomenological constants which can be checked in experimental data.

The main goal of this paper is to elaborate the model-framework
parametrization of effective
coupling constants of the extended chiral lagrangian which is suitable for the
description of the low-energy matrix elements of vector, axial-vector,
scalar and pseudoscalar currents\ci{GL} as well as of the matrix elements of
the pseudoscalar gluon density\ci{DW}.
On this way we find the new set of OZI rules. In particular,
one of them predicts the branching ratio of
the decays $\psi' \rightarrow  J/\psi + \pi$ or $\eta$ \ci{DW}.
\section{$SU(3)$-chiral lagrangian to the $p^4$-order}
\hspace*{3ex} The phenomenological chiral lagrangian for $SU(3)$ octet
of pseudoscalar meson fields was elaborated in the
$p^2 + p^4$-order \ci{We,GL} of CPTh with taking
into account of basic symmetries of the QCD. This lagrangian has the
following structure,
\be
S_{eff} = \int d^4x ({\cal L}_2  + {\cal L}_4) \,+\, S_{WZW},
\la{lag1}
\ee
The Weinberg lagrangian is given by
\be
{\cal L}_2 = \frac{F^2_0}4[< (D_{\mu}U)^{\dagger} D^{\mu}U>
+ < {\chi}^{\dagger}U + U^{\dagger}\chi >]. \la{lag2}
\ee
where $F_0 \simeq 88$MeV is a bare pion-decay constant,
$U$ is a $SU(3)$-chiral field describing pseudoscalar mesons,
$U U^{\dagger} = 1,\,$ det$U = 1$. The external sources are accumulated in
the covariant derivative,
$ D_{\mu}=\partial_{\mu} + [V_{\mu},*] + \{A_{\mu},*\} $ with vector,
$ V_{\mu}$ and axial-vector, $A_{\mu}$ fields and in the
complex density, $\chi = 2 B_0( s + i p)$ with scalar and pseudoscalar
fields. The constant $B_0$ is related to the scalar quark condensate,
the order parameter of the spontaneous chiral symmetry breaking in the QCD.

The Gasser-Leutwyler lagrangian contains ten basic low-energy constants $L_i$,
\ba
{\cal L}_4^{GL}&=&
L_1< (D_{\mu}U)^{\dagger}D^{\mu}U>^2 +
L_2< (D_{\mu}U)^{\dagger}D_{\nu}U>\!
< (D^{\mu}U)^{\dagger}D^{\nu}U> \nl
&&+ L_3< (D^{\mu}U)^{\dagger}D_{\mu}U(D^{\nu}U)^{\dagger}
D_{\nu}U> +
L_4< (D^{\mu}U)^{\dagger}D_{\mu}U>< {\chi}^{\dagger}U
+ U^{\dagger}\chi> \nl
&&+ L_5< (D^{\mu}U)^{\dagger}D_{\mu}U({\chi}^{\dagger}U
+ U^{\dagger}\chi)>
 + L_6< {\chi}^{\dagger}U + U^{\dagger}\chi >^2 \nl
&& + L_7< {\chi}^{\dagger}U - U^{\dagger}\chi >^2 +
L_8< {\chi}^{\dagger}U{\chi}^{\dagger}U +
U^{\dagger}\chi U^{\dagger}\chi >\nl
&&+ L_9< F_{\mu \nu}^R D^{\mu}U(D^{\nu}U)^{\dagger} + F_{\mu
\nu}^L(D^{\mu}U)^{\dagger}D^{\nu}U> + L_{10}< U^{\dagger}F_{\mu
\nu}^R U F^{L \mu \nu}>  \la{lag3}
\ea
where $F_{\mu \nu}^L={\partial}_{\mu}L_{\nu}-{\partial}_{\nu
 }L_{\mu}+[L_{\mu} , L_{\nu}]$ and
$ L_{\mu}=V_{\mu}+A_{\mu};\,  R_{\mu}=V_{\mu}-A_{\mu}$.
The Wess-Zumino-Witten action $S_{WZW}$  assembles so-called anomalous vertices
\ci{GL,Do}
and it is not displayed here (see the next Sec.).

The derivation of the chiral lagrangian from the QCD by
direct bosonization\ci{AA} or by bosonization of a quark model
\ci{Ra,Ru,Vo} can be described schematically as follows,
\be
\int\!{\cal D}G\,{\cal D}q{\cal D}\bar q
\exp (- S(\bar q,\,q,\,G;\,\,V, A, s, p))
= \int\! {\cal D}U \exp(- S_{eff}(U;\,\, V, A, s, p) \cdot {\cal R}.
\la{bos}
\ee
where the averaging over gluons, $G_{\mu}$, and/or quarks, $q, \bar q$
is approximately replaced by the averaging over chiral fields $U$ and
then the effective action is expanded in derivatives of chiral and external
fields.
The model design consists in the choice of model parameters
(the type and size of the momentum cutoff, the spectral asymmetry or the
dynamical mass etc.) and
in the assumption that the remainder ${\cal R} \simeq 1$ at low energies.

In the large-$N_c$ approach the main contribution
into the effective lagrangian ${\cal L}_4$ is parametrized
by nine structural constants, $I_k$
\ba
{\cal L}_4^{eff} &=&
I_1< D_{\mu}U(D_{\nu}U)^{\dagger}D^{\mu}U(D^{\nu}U)^{\dagger}> +
 I_2< D_{\mu}U(D_{\mu}U)^{\dagger}D^{\nu}U(D^{\nu}U)^{\dagger}> \nl
&&+ I_3 < (D_{\mu}^2U)^{\dagger}D_{\nu}^2U> +
I_4< (D_{\mu}\chi)^{\dagger}D^{\mu}U + D_{\mu}\chi
(D^{\mu}U)^{\dagger}> \nl
&&+ I_5 < D_{\mu}U(D^{\mu}U)^{\dagger}(\chi
U^{\dagger} + U{\chi}^{\dagger})>
+ I_6 < U{\chi}^{\dagger}U{\chi}^{\dagger} + \chi U^{\dagger}\chi
U^{\dagger}> \nl
&&+ I_7< {\chi}^{\dagger}U - U^{\dagger}\chi >^2\nl
&&+ I_8< F_{\mu \nu}^R D^{\mu}U(D^{\nu}U)^{\dagger} + F_{\mu
\nu}^L(D^{\mu}U)^{\dagger}D^{\nu}U> + I_9< U^{\dagger}F_{\mu
\nu}^R U F^{L \mu \nu}>
\la{mf}
\ea
The constants $I_1,\ldots,I_6, I_8, I_9$ arise from one-loop quark
diagrams in the soft-momentum expansion (or, equivalently, from the
large-mass expansion of heavy-meson lagrangians \ci{EG})
where all pole contributions
of $SU(3)$-pseudoscalar mesons are amputated. Thereby $I_k = O(N_c);\, k\not=
7$. The coefficient $I_7$ is essentially
saturated by the vacuum pseudoscalar
configurations \ci{Ve} and it is estimated in the next Sec.
The vertices $I_3,\, I_4$
induce the off-shell momentum dependence of the decay
constant of pseudoscalar mesons that
finds a firm explanation due to existence of highly excited
pseudoscalar mesons \ci{AM} ("radial excitations").

The relations between coefficients $L_i$ and $I_j$ are derived when one
imposes the equations of motion from the Weinberg lagrangian ${\cal L}_2$
following the scheme of CPTh,
\be
 (D_{\mu}^2U)^{\dagger}U-U^{\dagger}D_{\mu}^2U-{\chi}^{\dagger}U
 + U^{\dagger}\chi
=\frac{1}{3}< U^{\dagger}\chi -{\chi}^{\dagger}U>,
\la{Eq}
\ee
and the identities for $SU(3)$-chiral currents,
$A_{\mu} = U^{\dagger} \partial_{\mu}U$ are applied \ci{GL},
\be
<  A_{\mu}A_{\nu}A_{\mu}A_{\nu} > =
- 2 <  A_{\mu}^2 A_{\nu}^2 > + \frac{1}{2}
<  A_{\mu}^2 >^2 + <  A_{\mu}A_{\nu} >^2
\ee
They read,
\ba
2L_1&= & L_2= I_1,\quad L_3 =  I_2 + I_3 - 2I_1,
\quad L_4 = L_6 = 0,\quad L_5 = I_4 + I_5,\nl
L_7 &=& I_7 -\frac{1}{6} I_4 + \frac{1}{12} I_3,
\quad L_8 =  -\frac{1}{4} I_3 + \frac{1}{2} I_4 + I_6,\quad
L_{9} = I_8,\quad L_{10} =  I_9 .
\la{Zw0}
\ea
{}From these relations one can see that seven phenomenological constants
$L_i,\, i\not=2,4,6$ are independent in the large-$N_c$ approximation
and they are parametrized by nine model parameters $I_j$. Therefore
there exists the two-parameter family of models which yield the same
effective lagrangian for $SU(3)$-pseudoscalar mesons in external
$V, A, s, p$ fields. In particular one could
select out $I_3 = I_4 = 0$ as it is conventionally
accepted. On the other hand the latter ones are determined
by the mixing of light pseudoscalar mesons with their excitations
\ci{AM}.

We remark that the Kaplan-Manohar
freedom under the quark mass
reparametrization \ci{DW} is fixed here by the choice of $L_6 = 0$
compartible with the large-$N_c$ estimations.

\section{$U_A(1)$-extension of the chiral lagrangian}
\hspace*{3ex} Let us perform the extension of the $SU(3)$-chiral lagrangian
which describes the bosonization of flavor-singlet currents and of the
pseudoscalar gluon density. As it is well known \ci{Do,Ec} the
straigthforward $U(1)$-extension of the $SU(3)$-lagrangian describes
correctly the properties of pseudoscalar meson $\eta'$ if one takes
into account the strong vacuum effects due to nontrivial topological
susceptibility \ci{Ve}.

In order to describe the meson matrix elements of pseudoscalar gluon density
we follow \ci{DW} and supplement the QCD lagrangian with the
singlet pseudoscalar source of this density,
\be
{\cal L}_{\theta} = - \theta (x) G \widetilde
G;\quad\mbox{where}\quad G \widetilde G \equiv
 \frac{\alpha_s}{8\pi} G^a_{\mu \nu}\widetilde
G^{\mu \nu}_a . \la{an1}
\ee
After the $SU(3)$-bosonization in accordance
to (\ref{bos}) one arrives \ci{DW} to the
lagrangian (\ref{lag1}), (\ref{lag2}), (\ref{lag3}) with modified
$\widetilde\chi = \chi \exp(i \theta(x)/3)$ and to the additional structural
vertices $L_i,\, i=14,\ldots, 19$,
\ba
{\cal L}_4^{(D\theta )}&= &  iL_{14}D_{\mu}D^{\mu}\theta
 < {\widetilde \chi}^{\dagger} U - U^{\dagger} \widetilde\chi> +
 iL_{15}D_{\mu}\theta < (D^{\mu} \widetilde\chi)^{\dagger} U
 - U^{\dagger}(D^{\mu} \widetilde\chi )> \nl
 &+& L_{16} D_{\mu}\theta D^{\mu}\theta
  < D_{\nu}U(D^{\nu}U)^{\dagger}> +
 L_{17} D_{\mu}\theta D_{\nu}\theta
  < D^{\mu}U(D^{\nu}U)^{\dagger}> \nl
 &+& L_{18}D_{\mu}\theta D^{\mu}\theta
 <\widetilde\chi U^{\dagger} + U \widetilde\chi^{\dagger}>+
 iL_{19}D_{\mu}\theta < U(D^{\mu}U)^{\dagger} D_{\nu}U
 (D^{\nu}U)^{\dagger}>.
\la{DW}
\ea
Herein the equations of motion (\ref{Eq}) have been imposed.
In comparison to \ci{DW} the vertex with $L_{19}$ is added. It survives
in the chiral limit $m_q \rightarrow 0$ and may be important in the Skyrmion
physics.

Let us consider now the $SU(3)$-bosonization in the model framework
which brings the chiral lagrangian (\ref{DW}). It consists
typically of the following stages. At first, one extends
the $SU(3)$-bosonization to the $U(3)$ one and introduce the collective
pseudoscalar variables, $U(x) \rightarrow \widetilde U = U \exp(i \eta_0 /3)$.
Then after the bosonization of one quark-loop action
one reveals in general two relevant vertices
describing the dynamics of singlet pseudoscalar bound state at leading
order of the derivative expansion,
\be
{\cal L}_- (\eta_0) = \Biggl(\eta_0 + i \xi <  \chi^{\dagger}
\widetilde U - \widetilde U^{\dagger} \chi>\Biggr) G \widetilde G.
\la{an2}
\ee
Evidently $\xi = O(1)$ at large $N_c$.
When combining this functional with the vertex (\ref{an1}) we see that
the external source $\theta(x)$ can be reabsorbed into the field $\eta_0$
by shifting $\eta_0\rightarrow \eta_0 - \theta$. As a result
the essential part of the lagrangian (\ref{DW})
(except for the vertex $L_{14}$) is created directly from the model
framework one(\ref{mf}),
\ba
L_{15}&= & - 6 L_{18}  =
-\frac{2}{3}(I_4 + I_5)= -\frac{2}{3} L_5 = (- 0.9 \pm 0.3)\cdot 10^{-3} ;\nl
 L_{16}&= &  \frac{1}{2} L_{17}  = -\frac{1}{6} L_{19}
 =
 \frac{2}{9}(I_1 + I_2 + I_3)=  \frac{2}{9}(3L_2 + L_3) = (0 \pm 1.5)\cdot
 10^{-3},
 \la{Zw1}
\ea
after the appropriate
rotation $\widetilde U = U \exp[i( \eta_0  - \theta)/3]$, the redefinition
$\widetilde\chi = \chi \exp(i \theta(x)/3)$ and the elimination of
heavy singlet field $\eta_0$ in the soft-momentum limit.
The numerical estimations
for the coefficients $L_2,\, L_3,\, L_5$
in (\ref{Zw1}) are taken from \ci{Do}.

The next step is to obtain the effective vertices for $\eta_0$
which are substantially induced by the
QCD vacuum effects leading to the nontrivial correlator at zero momenta,
\be
M_0^4 = -\int d^4\!x <  0|T(G\widetilde G(x) G\widetilde
G(0))|0>_0 ,
\ee
in accordance to the common solution of $U(1)$-problem \ci{Ve}.

After averaging over gluons we derive the additional contribution into
the $U(3)$ meson lagrangian,
\be
\widetilde{\cal L} (\eta_0) = - \frac{M_0^4}{2}
\Biggl(\eta_0 + i \xi <  \chi^{\dagger}
\widetilde U - \widetilde U^{\dagger} \chi>\Biggr)^2
\ee
Since $M_0$ is not a soft chiral parameter one can apply the large-mass
reduction  for $\eta_0$ field and expand in powers of $1/M_0$. On this way
the $SU(3)$-chiral lagrangian is saturated in the vertices $L_7,\, L_{14}$,
\be
I_7 = - \frac{1}{2M_0^4}\Biggl(\frac{F_0^2}{12} - \xi M_0^4\Biggr)^2;\quad
L_7 = - \frac{F_0^4}{288 M_0^4} -\frac{1}{6} I_4 + \frac{1}{12} I_3
+ \frac{\xi F_0^2}{12} - \frac{\xi^2 M_0^4}{2} .
\la{Zw7}
\ee
where the first term in $L_7$ is of order $O(N_c^2), N_c =3$
and the next three ones are $O(N_c)$. Respectively, the
coefficient \be L_{14} = \frac{F_0^4}{72 M_0^4} -\frac{1}{3}
I_4 - \frac{2}{3} I_5 - \frac{\xi F_0^2}{6}  = - 4L_7 + O(N_c),
\la{Zw2}
\ee
is remarkably related to the structural constant $L_7$ in the main
large-$N_c$ order. The latter one is well estimated
from the $\eta,\,\eta'$-meson mass spectrum\ci{Do},
$L_7 = (- 0.4 \pm 0.2)\cdot 10^{-3}$.
Therefore it is expected that $L_{14} \simeq 1.6 \cdot 10^{-3}$.
This parameter
is involved into the description of the branching ratio of decays
$\psi'(3685) \rightarrow  J/\psi(3097) + \pi$ or $\eta$ \ci{DW}.
It is determined by the following matrix elements,
\be
\Biggl|\frac{<  0|G\widetilde G|\pi^0>}
{< 0|G\widetilde G|\eta >}\Biggr|^2 = (3.6 \pm 0.9) 10^{-2}
\ee
in the assumption that the multipole expansion is valid \ci{VZ}. It gives
the estimation, $L_{14} = (2.3 \pm 1.1)\cdot 10^{-3}$.
One can convince oneself that the Zweig rule prediction (\ref{Zw2}) is quite
satisfactory and can be thought of as the justification of multipole
expansion approach for the above decays.
\section{Conclusion}
\hspace*{3ex} We have established the relations between the model parameters
$I_j,\,j= 1,\ldots,9;\, M_0,\, \xi$ and
the phenomenological coupling constants
$L_i, \,i = 1,\ldots, 10, 14,\ldots,19$ which are typical for a wide class
of QCD-inspired quark models and can be verified in the experimental data.
It happens that the coupling constants $L_{14} \ldots L_{19}$ of the extended
chiral lagrangian are involved in a number of Zweig-type rules and can be
related to the structural constants $L_2, L_3, L_5, L_7$ in the main
large-$N_c$ order.

In the common approach to the parametrization of $\eta,\,\eta'$ meson
masses \ci{GL,Do,Ve} one neglects the constant $\xi \simeq 0$ (see,
however,\ci{Ni}). It holds
also for the
particular QCD-bosonization models \ci{AA}. In the assumption that
$\xi = 0$ and the susceptibility $M^4_0$ is known we can invert
the relations (\ref{Zw0}),(\ref{Zw1}),
(\ref{Zw7}),(\ref{Zw2}) and evaluate the model parameters from the
physical input,
\ba
I_1&=&L_2;\quad
I_2 = L_3 + 2L_2 - 4L_5 + \frac{F_0^4}{24M_0^4} - 12L_7 - 6L_{14};\nl
I_3&=&4L_5 - \frac{F_0^4}{24M_0^4} + 12L_7 + 6L_{14};\quad
I_4 = 3L_{14}+2L_5-\frac{F_0^4}{24M_0^4}; \nl
I_5&=&- 3L_{14} - L_5 + \frac{F_0^4}{24M_0^4};\;
I_6 = 3L_7+L_8 + \frac{F_0^4}{96M_0^4};\; I_7 = - \frac{F_0^4}{288M_0^4}.
\ea
We see that the tachyonic vertices with $I_3,\,I_4$ can be principally
determined at a given $M_0$. Thus the excited pseudoscalar
mesons \ci{AM} make influence on the low energy meson dynamics.

On the other hand in the Gauged NJL models \ci{Ra} one can find the vertex with
$\xi \not= 0$ though it is small ($ \xi << F_0^2/12 M_0^4$). When existing
such a vertex
modifies the mass formulae for $\eta, \eta'$ mesons that will be considered
elsewhere.  \\
\hfill

We are very grateful to the organizers  of the Workshop on Chiral Perturbation
Theory (NORDITA, Karreb{\ae}ksminde, Oct.1993)
for providing the opportunity to present our results
and especially to Prof. J. Bijnens for fruitful discussions and  financial
support. This paper is partially supported by the International Science
Foundation (G. Soros Foundation). A. A. is supported by the
Russian Grant Center for Natural Sciences.

\end{document}